\documentclass[aps,floats,showpacs,nofootinbib,floatfix]{revtex4}%
\usepackage{epsfig}
\usepackage{amsfonts}
\usepackage{amsmath}
\usepackage{graphicx}
\usepackage{amssymb}%
{\catcode`\|=\active \gdef\Braket#1{\left<\mathcode`\|"8000\let|\bravert
{#1}\right>}}
\def\bravert{\egroup\,\vrule\,\bgroup}

\begin{document}
\title{The eta-photon transition form factor}
\author{S. Noguera$^{a)}$, S. Scopetta$^{b)}$}
\email{Santiago.Noguera@uv.es, sergio.scopetta@pg.infn.it}
\affiliation{$a)$ Departamento de Fisica Teorica and Instituto de F\'{\i}sica Corpuscular,
Universidad de Valencia-CSIC, E-46100 Burjassot (Valencia), Spain. }
\affiliation{$b)$ Dipartimento di Fisica, Universit\`a di Perugia, and INFN, Sezione di
Perugia, via A. Pascoli, I-06100 Perugia, Italy.}
\date{\today }

\begin{abstract}
The eta-photon transition form factor is evaluated in a formalism based on a
phenomenological description at low values of the photon virtuality, and a
QCD-based description at high photon virtualities, matching at a scale
$Q_{0}^{2}$. The high photon virtuality description makes use of a
Distribution Amplitude calculated in the Nambu-Jona-Lasinio model with
Pauli-Villars regularization at the matching scale $Q_{0}^{2}$, and QCD
evolution from $Q_{0}^{2}$ to higher values of $Q^{2}$. A good description of
the available data is obtained. The analysis indicates that the recent data
from the BaBar collaboration on pion and eta transition form factor can be
well reproduced, if a small contribution {of higher twist is added to the 
dominant twist two contribution} at the matching scale
$Q_{0}^{2}$.

\end{abstract}

\maketitle

\section{Introduction}

Meson Distribution Amplitudes ($DA$) are fundamental theoretical ingredients
in the description of exclusive high energy processes. The pseudoscalar
transition form factors ($TFF$), $F_{\gamma\gamma^{\ast}P},$ describing the
process $P\rightarrow\gamma\gamma^{\ast},$ where $P$ is a pseudoscalar meson,
are directly connected with the $DA$s. Recently, the BaBar Collaboration has
provided new data at high virtuality for the pion and eta $TFF$ ($\pi TFF$ and
$\eta TFF$) \cite{:2009mc, :2011hk}. The implications of these results on $\pi
TFF$ in our understanding of the pion structure have been widely
discussed 
\cite{Radyushkin:2009zg,Polyakov:2009je,Mikhailov:2009sa,Dorokhov:2009zx,
Kochelev:2009nz,Noguera:2010fe,Arriola:2010aq,Leitner:2010nx,Kroll:2010bf,Wu:2010zc,Agaev:2010aq}%

In particular, these results have cast doubts on the behavior, as a function
of the light-cone momentum fraction $x$, of the pion distribution amplitude
($\pi DA$) $\phi_{\pi}(x)$ \cite{Efremov:1979qk,Chernyak:1981zz}, a quantity
for which some investigations have predicted a flat behavior, i.e., a constant
value for any $x$ \cite{Radyushkin:2009zg,Polyakov:2009je}, in good agreement
with the data of the form factor. These scenarios are compatible with QCD sum
rules \cite{Chernyak:1981zz} and lattice QCD
\cite{DelDebbio:2005bg,Braun:2006dg} calculations which provide values for the
second moment of the $\pi DA$ which are large compared to the asymptotic value
$6x(1-x)$. Several model calculations, such as the ones performed in the
Nambu-Jona-Lasinio (NJL)
\cite{RuizArriola:2002bp,Courtoy:2007vy,CourtoyThesis} or in the
\textquotedblright spectral\textquotedblright\ quark model
\cite{RuizArriola:2003bs} frameworks, give a constant $\pi DA$, i.e.
$\phi_{\pi}(x)=1$.

With the availability of data about the eta, it is important to analyze all
the proposed theoretical schemes. Some work in this direction has been already
done {{\cite{Kroll:2010bf,Xiao:2005af,
Agaev:2010zz,Wu:2011gf,Brodsky:2011yv,Klopot:2011qq,Klopot:2010ke}. 
In particular, in  Refs.  
\cite{Kroll:2010bf,Xiao:2005af,Agaev:2010zz,Wu:2011gf,
Brodsky:2011yv} 
the importance of the transverse momentum of the quarks is emphasized,
making use of some parametrization of the eta wave function.
In Refs. \cite{Klopot:2011qq,Klopot:2010ke}, the
TFFs of pseudoscalar mesons are considered by using a dispersive
representation of the axial anomaly, considering also the violation 
of factorization and possible higher twist corrections.
}}

The parton distributions, generalized parton distributions and distribution
amplitudes have been used as a test of hadron models. The procedure consists
of three ingredients: \textit{\ i}) the hadron model provides a low energy
description of the studied distribution; \textit{\ ii}) a high energy
description is obtained by QCD evolution, which needs an input at some low
scale $Q_{0}^{2}$; \textit{\ iii}) a matching condition between the two
descriptions at a scale $Q_{0}^{2}$ characterizing the separation between the
two regimes. This procedure has been useful in the study of nucleon parton
distributions \cite{Traini:1997jz,Scopetta:1997wk,Scopetta:1998sg} as well as
in that of pion distributions
\cite{Davidson:1994uv,Theussl:2002xp,Noguera:2005cc, Broniowski:2007si,
Courtoy:2008nf}.

In Ref. \cite{Noguera:2010fe}, a version of the previous program, but in a
rather model-independent formalism, has been used to calculate the $\pi TFF$.
An excellent description of experimental data has been obtained in the whole
range of virtuality. Summarizing, the evaluation of the $\pi TFF$ at high
$Q^{2}$ values in Ref. \cite{Noguera:2010fe} is based on the following
arguments:\textit{\ i}) chiral symmetry and soft pion theorems, which explain
that, at some point $Q_{0}^{2}$, the $\pi DA$ has a flat behavior, $\phi_{\pi
}(x,Q_{0}^{2})=1;$ \textit{ii}) applying QCD evolution to the $\pi DA$, one
can obtain the $\pi TFF$ at any $Q^{2}\gg Q_{0}^{2},$ \textit{iii}) for
$Q^{2}<Q_{0}^{2}$, the experimental parametrization of $F_{\gamma\gamma^{\ast
}\pi}(Q^{2})$ given in ref. \cite{Gronberg:1997fj} is assumed; \textit{iv})
for $Q^{2}>Q_{0}^{2}$ the $\pi TFF$ is given by its standard expression in
terms of the $\pi DA$ {modified }
in two directions, the quark propagator is {corrected }, as suggested by 
Radyushkin \cite{Radyushkin:2009zg},
and a term originated by other higher twist contributions is included.

This scheme, successful for the pion, requires further tests. The most natural
one consists in the evaluation of the same quantity for the other pseudoscalar
mesons. In this paper, the program is developed for the $\eta$ meson, for
which a few sets of data are available \cite{:2011hk,Gronberg:1997fj}. The
importance of the $\eta-\eta^{\prime}$ system for our understanding of the QCD
symmetries, and for their treatment in effective, low energy descriptions, is
well known (see, i.e., Ref. \cite{Feldmann:1999uf} and references therein). To
implement this program, the approach of Ref. \cite{Noguera:2010fe} has to be
complicated, and some hints have to be obtained within a specific model. In
particular, a generalized SU(3) Nambu-Jona-Lasinio model with Pauli-Villars
regularization, along the lines of Ref. \cite{Klevansky:1992qe}, will be used.

The NJL model is the most realistic model for the pseudoscalar mesons based on
a local quantum field theory built with quarks. It respects the realization of
chiral symmetry and gives a good description of the low energy physics of
pseudoscalar mesons. It allows to describe mesons in a field theoretical
framework treating them as bound states in a fully covariant manner
using the Bethe-Salpeter amplitude. In this way, the Lorentz covariance of the
problem is preserved.

The NJL model is a non-renormalizable field theory and therefore a cut-off
procedure has to be implemented. The Pauli-Villars regularization
procedure has been chosen because it respects the gauge symmetry of the
problem. The NJL model together with its regularization procedure is regarded
as an effective theory of QCD. In the chiral limit, it predicts for the pion
$\phi_{\pi}\left(  x\right)  =1,$ $0\leq x\leq1,$ in agreement with the model
independent study of the pion in Ref. \cite{Noguera:2010fe}. At this
point, due\ to the lack of fits in the $SU(3)$ NJL model with the
Pauli-Villars regularization , it is performed a new analysis of the
parameters of the model. It is interesting to notice that an early attempt to
use the NJL model, but in a $U(3)$ invariant scheme, has been presented in
Ref. \cite{Davidson:2001cc}, where the $\eta$ parton distribution has been evaluated.

The paper is organized as follows. In section II, the theoretical description
of the $\gamma\gamma^{\ast}\rightarrow P$ is reviewed, extracting the soft
(non-perturbative) part, to be described by the model. In section III, the
$\eta$ calculation in the NJL model is presented. In the following
section, numerical results are presented and discussed. The conclusions are
drawn in the last section. In the Appendix, a summary of the NJL model is
given, including the description of the fit which has been used.

\section{The $\gamma\gamma^{\ast}\rightarrow P$ process: theoretical
description.}

The subject of this study is the transition form factor, $F_{\gamma
\gamma^{\ast}P}$, i.e., the form factor for the coupling of a real photon and
a virtual photon to a pseudoscalar meson, $P$. The TFF is a very important
quantity in the QCD description of exclusive processes. In particular, it can
be used to obtain information on the shape of the meson $DA$
\cite{Efremov:1979qk,Lepage:1980fj,Chernyak:1983ej}. Experimentally, it has
been measured for the $\pi$, $\eta$ and $\eta^{\prime}$ mesons by the CELLO
\cite{Behrend:1990sr}, by the CLEO \cite{Gronberg:1997fj} and, recently, by
the BaBar \cite{:2009mc,:2011hk} collaborations. The latter results, for the
pion, have been found in disagreement with theoretical expectations.

In order to establish the proper formalism, in this section the theoretical
description of the $\gamma\left(  q_{1}\right)  \gamma^{\ast}\left(
q_{2}\right)  \rightarrow P\left(  k\right)  \,$ process is reviewed. From
general arguments it is well known that the transition amplitude of this
process can be written as (see, i.e., \cite{Itzykson:1980rh}):%
\begin{equation}
\left\langle P\left(  k\right)  \text{out}|\gamma\left(  q_{1},\varepsilon
_{1}\right)  ,\gamma^{\ast}\left(  q_{2},\varepsilon_{2}\right)
\text{in}\right\rangle =i\,\left(  2\pi\right)  ^{4}\delta^{4}\left(
q_{1}+q_{2}-k\right)  \,\mathcal{T}~, \label{F.01}%
\end{equation}
with%
\begin{equation}
\mathcal{T}=4\,\pi\,\alpha\,\varepsilon_{1\mu}\,\varepsilon_{2\nu}\,q_{1\rho
}\,q_{2\sigma}\,\varepsilon^{\mu\nu\rho\sigma}\,F_{\gamma\gamma^{\ast}P}~,
\label{F.02}%
\end{equation}
where $\alpha$ is the fine structure constant. On the other hand, applying the
reduction formalism of Lehmann, Symanzik and Zimmermann
\cite{Itzykson:1980rh}, this process is described by%
\begin{align}
&  \left\langle P\left(  k\right)  \text{out}|\gamma\left(  q_{1}%
,\varepsilon_{1}\right)  ,\gamma^{\ast}\left(  q_{2},\varepsilon_{2}\right)
\text{in}\right\rangle \nonumber\\
&  =-\left(  2\pi\right)  ^{4}\delta^{4}\left(  q_{1}+q_{2}-k\right)  \int
d^{4}z\,e^{-i(q_{1}-q_{2})z/2}\,\left\langle P\left(  k\right)  \text{out}%
\right\vert T\left(  \varepsilon_{1}\cdot j\left(  \frac{z}{2}\right)
\,\varepsilon_{2}\cdot j\left(  -\frac{z}{2}\right)  \right)  \left\vert
0\right\rangle ~, \label{F.03}%
\end{align}
where $j_{\mu}\left(  z\right)  =\sum_{j}e_{j}$ $\bar{\psi}_{j}\left(
z\right)  \gamma_{\mu}\psi_{j}\left(  z\right)  $ is the electromagnetic
current for the quarks. Evaluating the time ordered product at leading order
and after some algebra, one has%
\begin{align}
\mathcal{T}  &  =i\int d^{4}z\,e^{-i(q_{1}-q_{2})z/2}\,\int\frac{d^{4}%
t}{\left(  2\pi\right)  ^{4}}e^{-itz}\nonumber\\
&  \,\sum_{j}2\,i\,e_{j}^{2}\,\varepsilon_{1\mu}\,\varepsilon_{2\nu}\,\left(
-i\varepsilon^{\mu\alpha\nu\beta}t_{\alpha}\right)  \frac{1}{t^{2}-m_{j}%
^{2}+i\epsilon}\left\langle P\left(  k\right)  \text{out}\right\vert \bar
{\psi}_{j}\left(  \frac{z}{2}\right)  \gamma_{\beta}\gamma_{5}\psi_{j}\left(
-\frac{z}{2}\right)  \left\vert 0\right\rangle ~. \label{F.05}%
\end{align}
Defining%
\begin{equation}
\left\langle P\left(  k\right)  \text{out}\right\vert \bar{\psi}%
_{j\mathfrak{a}}\left(  \frac{z}{2}\right)  \psi_{j\mathfrak{b}}\left(
-\frac{z}{2}\right)  \left\vert 0\right\rangle =\int\frac{d^{4}\ell}{\left(
2\pi\right)  ^{4}}e^{-i\left(  \frac{k}{2}-\ell\right)  z}\,i\,\chi
_{\mathfrak{ab}}^{j}\left(  \ell\right)  ~, \label{F.07}%
\end{equation}
where $\mathfrak{a,b}$ are quadrispinor indices, one can write the amplitude
$\mathcal{T}$ as,%
\begin{equation}
\mathcal{T}=\varepsilon_{1\mu}\,\varepsilon_{2\nu}\,\varepsilon^{\mu\alpha
\nu\beta}\,I_{\alpha\beta}~, \label{F.10}%
\end{equation}
with%
\begin{equation}
I_{\alpha\beta}=-i\,\int\frac{d^{4}\ell}{\left(  2\pi\right)  ^{4}}2\,\left(
q_{1}-\ell\right)  _{\alpha}\,\sum_{j}\frac{1}{\left(  q_{1}-\ell\right)
^{2}-m_{j}^{2}+i\epsilon}\,e_{j}^{2}\,\left(  \gamma_{\beta}\gamma_{5}\right)
_{\mathfrak{ba}}\left(  \,i\,\chi_{\mathfrak{ab}}^{j}\left(  \ell\right)
\right)  ~. \label{F.12}%
\end{equation}
Since the symmetric part of $I_{\alpha\beta}$ doesn't give contribution to
$\mathcal{T}$, the attention is focused in the antisymmetric part. With the
momenta $q_{1}$ and $q_{2},$ one can build two antisymmetric tensors,
$q_{1\alpha}\,q_{2\beta}-q_{1\beta}\,q_{2\alpha}$ and $\varepsilon
_{\alpha\beta\rho\sigma}q_{1}^{\rho}q_{2}^{\sigma}.$ Nevertheless, there is
not enough structure in the integrand of Eq. (\ref{F.12}) to generate a tensor
like $\varepsilon_{\alpha\beta\rho\sigma}q_{1}^{\rho}q_{2}^{\sigma},$ at least
at the leading order. Therefore, the tensor structure of $I_{\alpha\beta}$ is
\begin{equation}
I_{\alpha\beta}=\frac{1}{2}\left(  q_{1\alpha}\,q_{2\beta}-q_{1\beta
}\,q_{2\alpha}\right)  \,I+\text{symmetric part}~. \label{F.15}%
\end{equation}
Turning back to the Eqs. (\ref{F.02}) and (\ref{F.10}), and using Eq.
(\ref{F.15}), one gets $F_{\gamma\gamma^{\ast}P}=-I.$ Contracting (\ref{F.15})
with $q_{1\alpha}\,q_{2\beta}-q_{1\beta}\,q_{2\alpha}$ and using the explicit
expression of $I_{\alpha\beta},$ Eq. (\ref{F.12}), yields%
\begin{align}
F_{\gamma\gamma^{\ast}P}  &  =-\frac{i}{\left(  q_{1}.q_{2}\right)  ^{2}%
}\,\int\frac{d^{4}\ell}{\left(  2\pi\right)  ^{4}}2\,\,\sum_{j}\frac
{1}{\left(  q_{1}-\ell\right)  ^{2}-m_{j}^{2}+i\epsilon}\,e_{j}^{2}%
\,\nonumber\\
&  \left[  q_{1}.\left(  q_{1}-\ell\right)  \left(  \rlap{$/$}q_{2}\gamma
_{5}\right)  _{\mathfrak{ba}}-q_{2}.\left(  q_{1}-\ell\right)  \left(
\rlap{$/$}q_{1}\gamma_{5}\right)  _{\mathfrak{ba}}\right]  \left(
\,i\,\chi_{\mathfrak{ab}}^{j}\left(  \ell\right)  \right)  ~. \label{F.18}%
\end{align}

This expression for the transition form factor is quite general. At this stage
the assumptions made are \textit{i}) the free quark propagator has been used
in going from Eq. (\ref{F.03}) to Eq. (\ref{F.05}) and \textit{ii}) the
corrections to the elctromagnetic vertex has not been considered. A more
general expression could be obtained by changing the free propagator, $\left(
\rlap{$/$}t-m+i\epsilon\right)  ^{-1}$, by the general one associated to a
dressed quark, $\left(  A\left(  t\right)  \,\rlap{$/$}t-B\left(  t\right)
+i\epsilon\right)  ^{-1}$, studied in actual lattice QCD calculations
\cite{Furui:2006ks} and including a term with the neglected\ structure,
$\varepsilon_{\alpha\beta\rho\sigma}q_{1}^{\rho}q_{2}^{\sigma}$.

Looking at the kinematics of the process, one can choose the reference frame
in such a way that the pion and photons four momenta are $k=\left(
E_{k},k_{x},0,k_{z}\right)  ,$ $q_{1}=\left(  E_{1},0,0,-E_{1}\right)  ,$
$q_{2}=\left(  E_{k}-E_{1},k_{x},0,E_{1}+k_{z}\right)  ,$ 
respectively. It is interesting to
express the quantities in terms of the light-front variables, $k^{\pm}=\left(
E_{k}\pm k_{z}\right)  /\sqrt{2},$ $\vec{k}_{\perp}=\left(  k_{x},0\right)  ,$
$q_{1}^{-}=\left(  Q^{2}+m_{P}^{2}\right)  /2k^{+},$ $q_{1}^{+}=0,$ $\vec
{q}_{1\perp}=0,$ $q_{2}^{+}=k^{+},$ $q_{2}^{-}=-\left(  Q^{2}-k_{\perp}%
^{2}\right)  /2k^{+},$ $\vec{q}_{2\perp}=\vec{k}_{\perp},$ where $Q^{2}%
=-q_{2}^{2}.$ In the limit of large $Q^{2}$, some of the quantities in Eq.
(\ref{F.18}) can be approximated by%
\begin{align}
\rlap{$/$}q_{1}  &  \simeq-\ \rlap{$/$}q_{2}\simeq\frac{Q^{2}}{2k^{+}}%
\gamma^{+}~,\label{F.20}\\
q_{1}.q_{2}  &  \simeq\frac{1}{2}Q^{2}~,\label{F.22}\\
\left(  q_{1}-\ell\right)  ^{2}-m_{j}^{2}+i\epsilon &  \simeq-\frac{\ell^{+}%
}{k^{+}}Q^{2}~, \label{F.25}%
\end{align}
giving for the transition form factor the expression%
\begin{equation}
F_{\gamma\gamma^{\ast}P}\simeq-\frac{1}{Q^{2}}i\,\int\frac{d^{4}\ell}{\left(
2\pi\right)  ^{4}}2\,\sum_{j}\frac{1}{\frac{\ell^{+}}{k^{+}}}\,e_{j}%
^{2}\,\left(  \frac{1}{k^{+}}\gamma^{+}\gamma_{5}\right)  _{\mathfrak{ba}%
}\left(  \,i\,\chi_{\mathfrak{ab}}^{j}\left(  \ell\right)  \right)  ~.
\label{F.26}%
\end{equation}

Finally, defining $x=\frac{\ell^{+}}{k^{+}}$, one arrives to the usual
expression
\begin{equation}
F_{\gamma\gamma^{\ast}P}(Q^{2})\simeq\frac{1}{Q^{2}}\int\frac{dx}{x}\Phi
_{P}\left(  x,Q^{2}\right)  ~, \label{F.28}%
\end{equation}
with%
\begin{align}
\Phi_{P}\left(  x,Q^{2}\right)   &  =-i\,\int\frac{d\ell^{-}d^{2}\ell_{\perp}%
}{\left(  2\pi\right)  ^{4}}2\,\sum_{j}\,e_{j}^{2}\,\left(  \gamma^{+}%
\gamma_{5}\right)  _{\mathfrak{ba}}\left(  \,i\,\chi_{\mathfrak{ab}}%
^{j}\left(  \ell\right)  \right) \nonumber\\
&  =-\,2\,i\,\int\frac{dz^{-}}{2\pi}\,e^{iz^{-}k^{+}\left(  x-\frac{1}%
{2}\right)  }\left.  \left\langle P\left(  k\right)  \text{out}\right\vert
\sum_{j}e_{j}^{2}\bar{\psi}_{j}\left(  \frac{z}{2}\right)  \gamma^{+}%
\gamma_{5}\psi_{j}\left(  -\frac{z}{2}\right)  \left\vert 0\right\rangle
\right\vert _{z^{+}=0,\vec{z}_{\perp}=0}\ \ \ . \label{F.30}%
\end{align}
As will be discussed later, in Eq. (\ref{F.28}), besides the explicit $Q^{2}$
dependence, also an implicit one appears, through the QCD evolution of
$\Phi_{P}\left(  x,Q^{2}\right)  $. In the SU(3) formalism, the quark operator
has the form%
\begin{equation}
\sum_{j}e_{j}^{2}\bar{\psi}_{j}\left(  \frac{z}{2}\right)  \gamma^{+}%
\gamma_{5}\psi_{j}\left(  -\frac{z}{2}\right)  =\bar{\psi}\left(  \frac{z}%
{2}\right)  \gamma^{+}\gamma_{5}\mathcal{O}\psi\left(  -\frac{z}{2}\right)  ~,
\label{F.32}%
\end{equation}
with $\mathcal{O}=2\,\lambda^{0}/(3\sqrt{6})+\lambda^{3}/6+\lambda^{8}%
/(6\sqrt{3})$ where $\lambda^{a}$ are the SU(3) generators. In the present
case it is more interesting to use the flavor basis in describing the $\eta$
particle (see the Appendix). In this basis, one has $\mathcal{O}%
=5\,\lambda^{q}/18+\lambda^{s}/(9\sqrt{2})+\lambda^{3}/6$. As usual, the $DA$
of $P$ in the flavor basis is defined as
\begin{equation}
if_{P}^{j}\phi_{P}^{j}(x,Q^{2})=-\int\frac{dz^{-}}{2\pi}\,e^{iz^{-}%
k^{+}\left(  x-\frac{1}{2}\right)  }\left.  \left\langle P\left(  k\right)
\right\vert \bar{\psi}_{j}\left(  \frac{z}{2}\right)  \gamma^{+}\gamma
_{5}\frac{\lambda^{j}}{\sqrt{2}}\psi_{j}\left(  -\frac{z}{2}\right)
\left\vert 0\right\rangle \right\vert _{z^{+}=0,\vec{z}_{\perp}=0}%
~,\ \ \ \label{fl_basis}%
\end{equation}
with $j=3,q,s$. This yields
\begin{equation}
\Phi_{P}(x,Q^{2})=\frac{\sqrt{2}}{3}f_{P}^{\,3}\phi_{P}^{3}(x,Q^{2})+\frac
{5}{9}\sqrt{2}f_{P}^{\,q}\phi_{P}^{q}(x,Q^{2})+\frac{2}{9}f_{P}%
^{\,s}\phi_{P}^{s}(x,Q^{2})~. \label{phip}%
\end{equation}
In the pion case, this equation corresponds to $\Phi_{\pi}\left(  x\right)
=\sqrt{2}\,f_{\pi}\,\phi_{\pi}\left(  x\right)  /3,$ where $\phi_{\pi}\left(
x\right)  $ is the $\pi DA$ and $f_{\pi}$= 0.131 GeV is the pion decay constant.

One should notice that, in going from Eq. (\ref{F.01}) to the final result Eq.
(\ref{F.28}), a few approximations have been done: 
the free expression has been used for the quark propagator, with the
additional simplification given by Eq. (\ref{F.25}); besides, the
approximations Eqs. (\ref{F.20}) and (\ref{F.22}) have been applied
in the numerator of Eq. (\ref{F.18}) and a new tensor structure in
$I_{\alpha\beta}$ has been neglected. Some of these corrections have a
kinematic character, while others are certainly dynamical.
Both type of corrections imply the presence of higher twist distribution
amplitudes.

In Ref. \cite{Noguera:2010fe} it has been argued that the approximations,
leading to the simple expression Eq. (\ref{F.28}) for the transition form
factor, are too crude to explain the BaBar experimental data, and corrections
at the next order in the $Q^{-2}$ expansion have been added. The simplest way
to implement these corrections is to start from the following expression:%
\begin{equation}
Q^{2}F_{\gamma\gamma^{\ast}P}(Q^{2})=\int_{0}^{1}\frac{dx}{x+\frac{M^{2}%
}{Q^{2}}}\Phi_{P}(x,Q^{2})+\frac{C_{3}}{Q^{2}}~. \label{tff_R_T3}%
\end{equation}
The mass $M$ in Eq. (\ref{tff_R_T3}) was introduced by Radyushkin
\cite{Radyushkin:2009zg}, to cure the divergence of the integrand in Eq.
(\ref{F.28}), occurring when a DA $\Phi_{P}\left(  x,Q_{0}^{2}\right)  $, not
vanishing at $x=0,1$, is used. This was justified as a consequence of the
existence of some transverse component in the quark momentum. As it has been
shown in the previous section, $M$ contains not only effects associated to the
mean transverse momentum, but also the ones associated to the constituent
quark masses, among others. In Ref. \cite{Noguera:2010fe} it has been shown
that it is necessary to introduce the $C_{3}$-dependent term in Eq.
(\ref{tff_R_T3}), otherwise the data cannot be well described around the
region of $Q^{2}=10-20\operatorname{GeV}^{2}$. The inclusion of this term has
been thoroughly motivated in this section, where it has been shown that the
perturbative approach leading to Eq. (\ref{F.28}) is correct only for high
enough values of the virtuality. { We call the $C_{3}$ term as
``the higher twist term''},
although it is clear that also the mass term, $M,$ is of the same order.

\section{The $\eta$ transition form factor}

In this section, we evaluate the $\eta TFF$. To this aim, according
to Eq. (\ref{tff_R_T3}), in calculating $F_{\gamma\gamma^{\ast}\eta}(Q^{2})$
\thinspace\ the $\eta DA$ is needed. From Eq.(\ref{phip}), the $\eta DA$ is
expressed by
\begin{equation}
\Phi_{\eta}\left(  x,Q^{2}\right)  =\frac{5}{9}\sqrt{2}f_{\eta}^{\,q}%
\,\phi_{\eta}^{q}\left(  x,Q^{2}\right)  +\frac{2}{9}\,f_{\eta}^{\,s}%
\,\phi_{\eta}^{s}\left(  x,Q^{2}\right)  ~. \label{eta.01}%
\end{equation}
In the calculation, we use the following values of the $\eta$ weak
decay constants
\begin{align}
f_{\eta}^{q}  &  =(0.828\pm0.019)f_{\pi}~,\nonumber\\
f_{\eta}^{s}  &  =(-0.848\pm0.042)f_{\pi}~, \label{feta}%
\end{align}
with $f_{\pi}=131%
\operatorname{MeV}%
,$ obtained in the phenomenological study of Ref. \cite{Feldmann:1999uf}.

Now,it is necessary to calculate the $\eta DA$ at some initial scale
$Q_{0}^{2}$ within a model. To this end, we have obtained the
DAs corresponding to the $q$ and $s$ flavors within the Nambu-Jona
Lasinio (NJL) model, which has a long tradition of successful
predictions of meson parton structure \cite{Davidson:1994uv, Theussl:2002xp,
Noguera:2005cc, Broniowski:2007si,Courtoy:2008nf}. In particular, we
use in the present calculation the three quark flavor version of the
model with the Pauli-Villars regularization \cite{Klevansky:1992qe,
Bernard:1988wi}. A brief summary of the model and of the regularization
procedure is given in the Appendix.

In the NJL model, mesons are described through Bethe-Salpeter amplitudes. For
the $\eta$ meson one has:%
\begin{align}
\left\langle \eta\left(  k\right)  \right\vert \psi_{\beta}\left(
x_{2}\right)  \bar{\psi}_{\alpha}\left(  x_{1}\right)  \left\vert
0\right\rangle  &  =e^{ik(x_{1}+x_{2})/2}\int\frac{d^{4}\ell}{\left(
2\pi\right)  ^{4}}e^{i\ell\left(  x_{1}-x_{2}\right)  }\nonumber\\
&  \left[  iS_{F}\left(  \ell-\frac{k}{2}\right)  ig_{\eta qq}\left(  \cos
\phi\,\lambda^{q}-\sin\phi\,\lambda^{s}\right)  i\gamma_{5}iS_{F}\left(
\ell+\frac{k}{2}\right)  \right]  _{\beta\alpha}%
\end{align}
where the index $\alpha,\beta$ stand for color, flavour and
quadrispinor index. Inserting this expression in Eq. (\ref{fl_basis})
one obtains%
\begin{align}
if_{\eta}^{\,j}\,\phi_{\eta}^{j}\left(  x\right)   &  =\sqrt{2}g_{\eta
qq}\left(  \cos\phi\,\delta_{j,q}-\sin\phi\,\delta_{j,s}\right)
N_{c}\nonumber\\
&  \int\frac{d^{4}\ell}{\left(  2\pi\right)  ^{4}}\delta\left(  \ell^{+}%
-k^{+}\left(  x-\frac{1}{2}\right)  \right)  tr\left(  S_{F}\left(  \ell
-\frac{k}{2}\right)  \gamma_{5}S_{F}\left(  \ell+\frac{k}{2}\right)
\gamma^{+}\gamma_{5}\right)  \ \ .
\end{align}
Evaluating the trace and using Eq. (\ref{Af}) of the Appendix, one has
\begin{equation}
\phi_{\eta}^{q(s)}(x,Q_{0}^{2})={\frac{1}{I_{2}(m_{q(s)},m_{q(s)},m_{\eta}%
^{2})}}\tilde{I}_{\eta,q(s)}(x,m_{q(s)},m_{\eta}^{2})~,\label{eta.05}%
\end{equation}
where $\tilde{I}_{2}\left(  x,m_{i},m_{P}^{2}\right)  $ is given in Eq.
(\ref{A.28}) and $I_{2}(m_{q(s)},m_{q(s)},m_{\eta}^{2})~$ is the
two-propagator integral defined in Eq. (\ref{A.02}). The flavor $DA$ defined
in Eq. (\ref{eta.05}) satisfies the normalization condition:%
\begin{equation}
\int dx\phi_{\eta}^{q(s)}(x,Q_{0}^{2})=1~.\label{phinorm}%
\end{equation}%

\begin{figure}
[ptb]
\begin{center}
\includegraphics[
height=4.2125in,
width=3.0986in
]%
{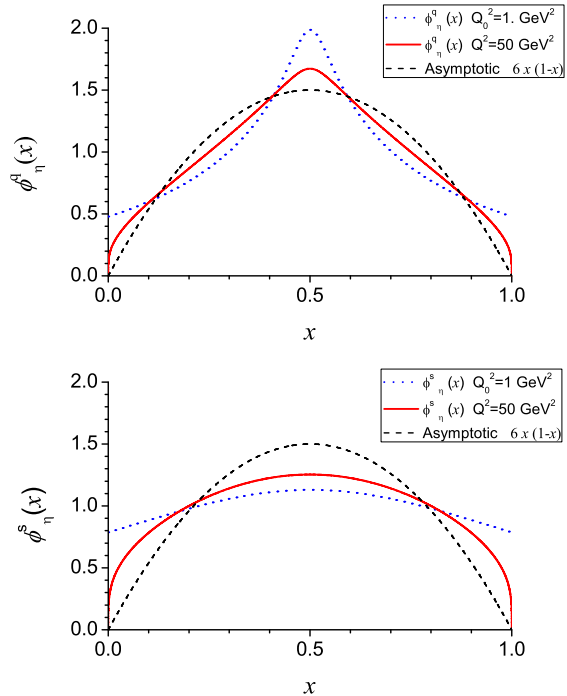}%
\caption{ The $DA$ for the $q$ (upper panel) and $s$ (lower panel) flavor in
the $\eta$ meson, at the initial scale $Q_{0}^{2}$ = 1 GeV$^{2}$ (dotted line)
and after evolution to the scale $Q^{2}$ = 50 GeV$^{2}$ (full line). The
asymptotic behavior is also shown for comparison (dashed line).}%
\label{Q01}%
\end{center}
\end{figure}

For $Q^{2}>Q_{0}^{2}$, the $\eta TFF$ is obtained through QCD evolution
\cite{Efremov:1979qk,Lepage:1979zb,Mueller:1994cn}. The $\eta DA$ can be
expressed in terms of the Gegenbauer polynomials, 
\begin{equation}
\phi_{\eta}^{q,s}(x,Q^{2})=6\,x\,(1-x)\sum_{n(even)=0}^{\infty}a_{n}%
^{q,s}\,C_{n}^{3/2}(2x-1)\left(  \frac{\log\frac{Q^{2}}{\Lambda_{QCD}^{2}}%
}{\log\frac{Q_{0}^{2}}{\Lambda_{QCD}^{2}}}\right)  ^{-\gamma_{n}%
},\label{pDA0eta}%
\end{equation}
where $\gamma_{n}$ is the anomalous dimension%
\begin{equation}
\gamma_{n}=\frac{C_{F}}{\beta}\left(  1+4\sum_{k=2}^{n+1}\frac{1}{k}-\frac
{2}{(n+1)(n+2)}\right)  ,\label{dima}%
\end{equation}
$\beta=\frac{11N_{C}}{3}-\frac{2N_{f}}{3}$ is the beta function to lowest
order and $C_{F}=\frac{N_{C}^{2}-1}{2N_{C}}.$ If the $a_{n}^{q,s}$
coefficients are known, using Eq. (\ref{pDA0eta}) in Eq. (\ref{eta.01}) and
(\ref{tff_R_T3}), the $\eta TFF$ is obtained for any $Q^{2}\gg\Lambda
_{QCD}^{2}.$ Once $\phi_{\eta}^{q,s}\left(  x,Q_{0}^{2}\right)  $ are known at
a given scale $Q_{0}^{2}$, the $a_{n}^{q,s}$ coefficients are obtained using
the orthogonality property of the Gegenbauer polynomials%
\begin{equation}
a_{n}^{q,s}=\frac{2}{3}\frac{2n+3}{\left(  n+1\right)  \left(  n+2\right)
}\int_{0}^{1}dx\,C_{n}^{3/2}(2x-1)\,\phi_{\eta}^{q,s}\left(  x,Q_{0}%
^{2}\right)  ~.\label{pDA1eta}%
\end{equation}

In this scheme, the $\eta$ meson cannot couple to two gluons. This should not
be a serious drawback of the approach, having the $\eta$ essentially an octet
character under $SU(3)$ transformations.

We fix now the values of $Q_{0}$, $C_{3}$ and %
$M.$ The $Q_{0}$ scale is closely related to the choice of
the $\Lambda_{QCD}$ value. We fix a scale of $Q_{0}=1$ GeV,
together with $\Lambda_{QCD}=0.226%
\operatorname{GeV}%
,$ in analogy with the previous analysis \cite{Noguera:2010fe}. A
natural condition to be satisfied is continuity between the low virtuality
description of the $\eta TFF$ and the high virtuality description,
provided by Eq (\ref{tff_R_T3}). To minimize the model dependence, we use the
parameterization of the CLEO collaboration \cite{Gronberg:1997fj} for the
description of the TFF in the LV region:%
\begin{equation}
F_{\gamma\gamma^{\ast}\eta}^{LV}\left(  Q^{2}\right)  =\left[  {\frac
{64\pi\Gamma(\eta\rightarrow\gamma\gamma)}{(4\pi\alpha)^{2}m_{\eta}^{3}}%
}\right]  {\frac{1}{1-{\frac{t}{\Lambda_{\eta}^{2}}}}}=\frac{F\left(
0\right)  }{1+\frac{Q^{2}}{\Lambda_{\eta}^{2}}}\,, \label{EXPeta}%
\end{equation}
with $F\left(  0\right)  =0.272\pm0.007%
\operatorname{GeV}%
^{-1},$ obtained using $\Gamma(\eta\rightarrow\gamma\gamma
)=0.510\pm0.026$ 10$^{-3}$ MeV as given by the Particle Data Group
\cite{Nakamura:2010zzi} together with $m_{\eta}=547.85$ MeV, and
$\Lambda_{\eta}=774\pm29$ MeV \cite{Gronberg:1997fj}.

The value of the mass $M$ can be obtained equating the $\eta TFF$
given by Eq. (\ref{tff_R_T3}) at $Q^{2}=Q_{0}^{2}$, using as $\Phi_{\eta
}(x,Q_{0}^2)$ the one provided by the NJL model, to the value given, at the same
scale, by the monopole parametrization Eq. (\ref{tff_R_T3}),
\begin{equation}
Q_{0}^{2}F_{\gamma\gamma^{\ast}\eta}^{LV}\left(  Q_{0}^{2}\right)
=\frac{Q_{0}^{2}F\left(  0\right)  }{1+\frac{Q_{0}^{2}}{\Lambda_{\eta}^{2}}%
}=\int_{0}^{1}\frac{dx}{x+\frac{M^{2}}{Q_{0}^{2}}}\Phi_{\eta}(x,Q_{0}%
^{2})+\frac{C_{3}}{Q_{0}^{2}}. \label{conteta}%
\end{equation}

Finally, the only unknown is $C_{3}$, for which several reasonable values have
been used, as discussed in the following section.%

\begin{figure}
[ptb]
\begin{center}
\includegraphics[
height=3.6244in,
width=5.1742in
]%
{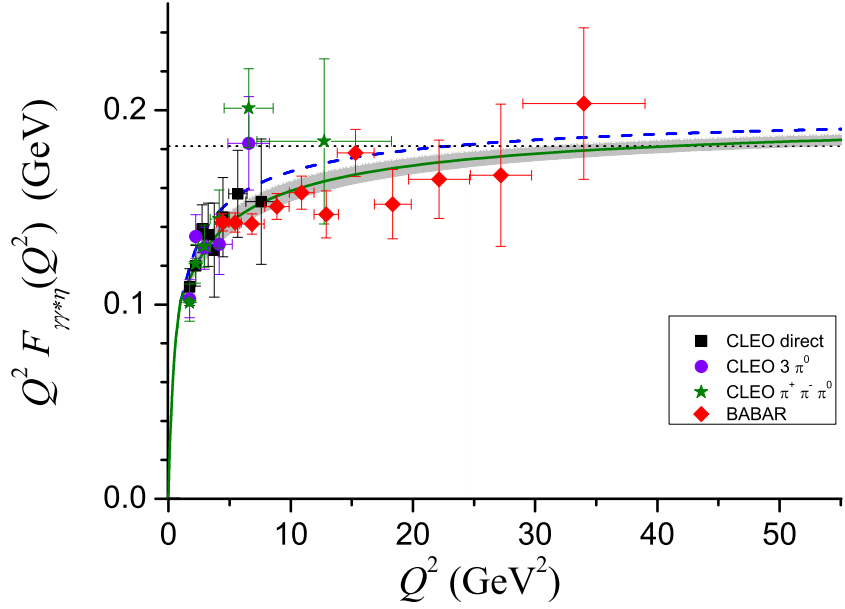}%
\caption{ Calculation of the transition form factor via the $\eta DA$ with
$M=0.557\operatorname{GeV}$, $\ C_{3}=2.04\,10^{-2}\operatorname{GeV}^{3}$ and
using $Q_{0}^{2}=1\operatorname{GeV}^{2}$ (full-line) compared with
the available experimental data \cite{:2011hk,Gronberg:1997fj}. The
gray region describes the indeterminacy on $C_{3}$. The dashed line
represents the result obtained taking $C_{3}=0$. The dotted line corresponds
to the asymptotic value for the form factor. }%
\label{Q01_0}%
\end{center}
\end{figure}

\begin{figure}
[ptb]
\begin{center}
\includegraphics[
height=3.6244in,
width=5.1742in
]%
{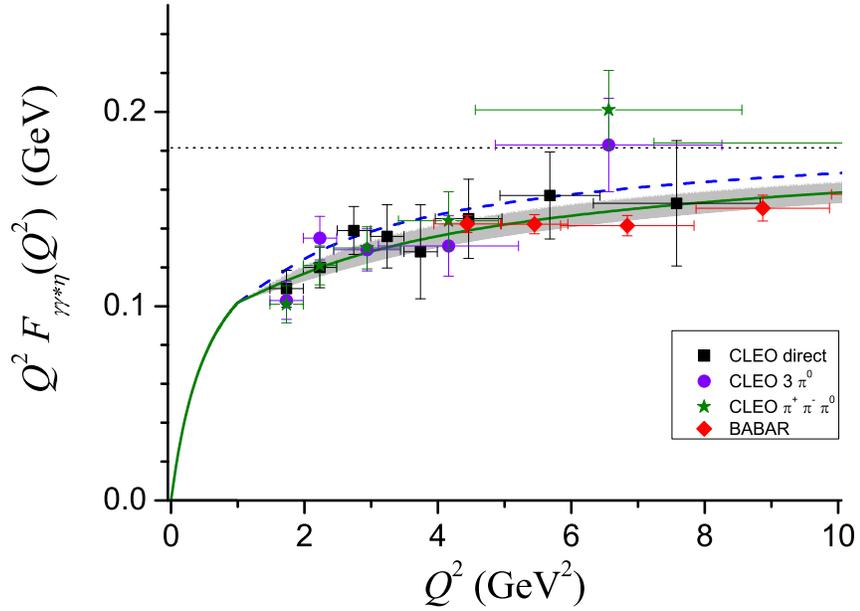}%
\caption{The same as in Fig \ref{Q01_0}, but in the low virtuality region}%
\label{Q01_1}%
\end{center}
\end{figure}

\section{Results and discussion}

\begin{table}[ptb]
\centering
\begin{tabular}
[c]{|l|c|c|c|c|c|c|c|c|c|}\hline
& $\mu_{s}$ & $m_{s}$ & $\left\langle \bar{s}s\right\rangle ^{1/3}$ & $m_{K}$
& $m_{\eta}$ & $m_{\eta^{\prime}}$ & $f_{K}/f_{\pi}$ & $f_{\eta}^{q}/f_{\pi}$
& $f_{\eta}^{s}/f_{\pi}$\\\hline
Set I & \ 171.2\  & \ 430\  & $\ -186\ $ & \ 497\  & \ 541\  & \ 1157\  &
\ \ 1.07 & \ \ 0.603 \ \ \  & \ $\ -0.664$\ \ \ \\
Set II\ \  & \ 184.2\  & \ 435\  & $-184$ & 515 & 554 & 1148 & \ \ 1.07 &
\ 0.832 \  & $\ -0.840 \ $\ \\
Exp. &  &  & $-194$ & 495 & 548 & 958 & \ 1.18 & \ 0.828\  & $-0.848$\\\hline
\end{tabular}
\caption{We show results for two different parametrizations of the
NJL model for several physical quantities, together with their experimental or
phenomenological values. Explicit expressions for these are reported
in the Appendix. The masses and the quark condensate are given in MeV.}%
\label{TableI}%
\end{table}
We present now our results of the calculation of the $\eta
TFF$. The starting point is the $\eta DA$, evaluated at the low energy scale
of the model. According to Eq. (\ref{eta.05}), all we need is the value of the
$\eta$ mass, the quark masses and the regularization parameter $\Lambda$.
We use for the $\eta$ mass the experimental value, $m_{\eta}=548%
\operatorname{MeV}%
.$ The quark masses $m_{q(s)}$ and the regularization parameter $\Lambda$ have
to be fixed within the NJL model. It is important to work in the Pauli-Villars
regularization scheme, in order to preserve gauge invariance. Unfortunately,
to our knowledge, all the available fits
for the NJL model in $SU(3)$ are done within the cutoff regularization scheme.
The only exception is the paper by Bernard and Vautherin \cite{Bernard:1988wi}%
, where anyway an approximate expression for the $I_{2}(m_{i},m_{j,}q^{2})$
integral is used. Therefore, we have performed a new fit of the model
parameters. The SU(3) NJL model gives a very good description for the meson
properties \cite{Klimt:1989pm}, but one has to be careful, since it does not
include confinement. To avoid problems, we impose a value of
$m_{u}>m_{\eta}/2.$ The details of the model (whose Lagrangian is given by Eq.
(\ref{A.01})) are given in the Appendix. Here it is worth to recall only that
the model has five parameters, which can be chosen as the current quark
masses, $\mu_{u}$ and $\mu_{s}$, the dressed quark masses, $m_{u}$ and
$m_{s},$ and the cutoff parameter, $\Lambda.$ { Our} strategy for the fits has
been: \textit{i}) the $m_{u}$ mass has been fixed to $275%
\operatorname{MeV}%
;$ \textit{ii}) $\mu_{u}$ and $\Lambda$ are determined by fitting $m_{\pi}$
and $f_{\pi};$ \textit{iii}) $\mu_{s}$ and $m_{s}$ are chosen looking for an
overall good description of the strange sector. In table \ref{TableI} two
different sets of the relevant quantities, obtained by the above described
fitting procedure, are reported. It is seen that their experimental values are
reproduced very well. In the Set I, we have imposed the additional
condition for the resulting eta mass: $m_{\eta}\leq2m_{u}.$ In this set, a
very good description of masses in the strange sector is obtained, but paying
the price of a worse description of $f_{\eta}^{\,q,s}.$ In Set II, the
description of the masses is slightly worse, but the $f_{\eta}^{\,q,s}$ are
very well reproduced. One should notice that, in going from $SU(2)$ to $SU(3)$
within the NJL model, the number of parameters moves from 3 to 5. One has
therefore at hand two more parameters for explaining three new masses, three
new decay constants and a new quark condensate. Actually, for calculating the
$\eta DA$ only $m_{u}$ and $\Lambda$, which have the same value in both sets,
are needed, together with $m_{s},$ which changes from 430 MeV in Set I to 435
MeV in Set II. With respect to these quantities, the predictions of the model
are therefore reasonably stable. The results for the $\eta DA$ are
presented for the following values of the parameters: $\Lambda=740%
\operatorname{MeV}%
,$ $m_{u}=275%
\operatorname{MeV}%
$ and $m_{s}=435%
\operatorname{MeV}%
.$ It may be useful to reiterate that the obtained values of  
$f_{\eta}^{\,q,s}$ are not relevant for the present calculation,
because we used the experimental ones.

In Fig. \ref{Q01}, the $\eta DA$ are shown.  We observe that
$\phi_{\eta}^{q}\left(  x,Q_{0}^{2}\right)  $ is peaked around the
central point $x=0.5$ while $\phi_{\eta}^{s}\left(  x\right)  $
is relatively flat. This is a consequence of the masses of quarks $u$
and $d$, which are close to half the mass of the eta, while it is not the case
for the mass of the strange quark. What is clearly seen is the following
reasonable feature: the less bound is a system, the more narrow is its DA
around the point $x=0.5$. 

{ Our $DAs$ have an infinite expansion in terms of 
the Gegenbauer polynomials.} The firsts coefficients
$a_{\eta}^{q,s}$, defined in Eq. (\ref{pDA1eta}), at $Q_{0}^{2}=1%
\operatorname{GeV}%
^{2}$ are:
\begin{align}
a_{2}^{q} &  =0.134\ \ \ \ \ \ \ \ \ \ \ \ \ \ a_{2}^{s}=0.377\nonumber\\
a_{4}^{q} &  =0.352\ \ \ \ \ \ \ \ \ \ \ \ \ \ a_{4}^{s}=0.245\label{apeque}%
\end{align}
The $a_{\eta}^{s}$ coefficients are close to the values predicted by a flat
distribution. On the other hand side, we observe that $a_{2}^{q}<a_{4}^{q},$
at variance with what is commonly used in the field. This feature is due to
the narrow structure of $\phi_{\eta}^{q}\left(  x\right)  .$ We can
compare our results with the values used by other authors. The
$a_{2}^{q}$ and $a_{2}^{s}$ coefficients are to be compared to the
parameter $B=0.3$ used in Ref. \cite{Wu:2011gf}. 

In Ref. \cite{Kroll:2010bf} the values $a_{2}^{1}=-0.06\pm0.06,$ $a_{2}^{8}%
=-0.07\pm0.04$ and ${a}_{2}^{\pi}{=0.22\pm0.06}$ are given, but at $Q^{2}=4%
\operatorname{GeV}%
^{2}.$ From Eq. (\ref{apeque}) and using $\phi_{\eta}^{1}=(5\,\phi_{\eta}%
^{q}+\phi_{\eta}^{s})/6$ and $\phi_{\eta}^{8}=(5\,\phi_{\eta}^{q}%
-2\,\phi_{\eta}^{s})/3$ we obtain%
\begin{equation}
a_{2}^{1}\left(  1%
\operatorname{GeV}%
^{2}\right)  =0.17\ \ \ \ \ \ \ \ \ \ \ a_{2}^{8}\left(  1%
\operatorname{GeV}%
^{2}\right)  =-0.028\ \ .\label{apeque2}%
\end{equation}
Evolving these results we have%
\begin{equation}
a_{2}^{1}\left(  4%
\operatorname{GeV}%
^{2}\right)  =0.14\ \ \ \ \ \ \ \ \ \ \ a_{2}^{8}\left(  4%
\operatorname{GeV}%
^{2}\right)  =-0.022~~.
\end{equation}
The value for $a_{2}^{8}$ is therefore consistent with that used in Ref.
\cite{Kroll:2010bf}, while some difference is found for the { value of} 
$a_{2}^{1}$. { In the pion case,}
if we use a flat distribution at $Q_{0}^{2}=1%
\operatorname{GeV}%
^{2},$ a value $a_{2}^{\pi}\left(  4%
\operatorname{GeV}%
^{2}\right)  =0.31$ is obtained. In Ref. \cite{Brodsky:2011yv} it has been
noted that the values for these parameters found in \cite{Kroll:2010bf}
suggest a very large SU(3) breaking between the $DA$ of the $\pi^{0}$ and
the one of $\eta^{8},$ and a very little U(1) symmetry breaking between
$\eta^{8}$ and $\eta^{1}.$ Our results show the same structure of
those of Ref \cite{Kroll:2010bf}, at least for $a_{2}^{8}$ and for the big
difference between $a_{2}^{\pi}${ and }$a_{2}^{8}.$ The origin of this
difference is in the small value of $a_{2}^{q}$ due to the narrow structure
of $\phi_{\eta}^{q}(x)$, originated by the fact that $m_{u}$ is close to
$m_{\eta}/2.$ At the same time, a small value of $a_{2}^{q}$ explains a
small value for $a_{2}^{1}.$ One should remember that the present scheme
reproduces the SU(3)$_{F}$ and U(3)$_{F}$ symmetry breaking in the
pseudoscalar meson sector.

The results of Eq. (\ref{apeque2}) are also in good agreement with
those from \cite{Agaev:2010zz,Agaev:2003kb}. In the latter references,
the authors give
their results for $Q_{0}^{2}=1%
\operatorname{GeV}%
^{2}$ in terms of the quantities
$B_{2}^{q}$ and $B_{2}^{g},$%
defined in \cite{Agaev:2003kb} and related to
our expressions { as follows }: 
$a_{2}^{1}\left(  1%
\operatorname{GeV}%
^{2}\right)  =-B_{2}^{q}+B_{2}^{g}/102$ and $a_{2}^{8}\left(  1%
\operatorname{GeV}%
^{2}\right)  =-B_{2}^{q}.$
{ Using the numerical results of } 
Ref \cite{Agaev:2010zz}, we
have $a_{2}^{1}\left(  1%
\operatorname{GeV}%
^{2}\right)  =0.149\pm0.048$ and $a_{2}^{8}\left(  1%
\operatorname{GeV}%
^{2}\right)  =-0.0425\pm0.0175,$ to be compared to our results, 
Eq. (\ref{apeque2}). In these papers, the coupling of a two gluon 
state to the singlet $\bar{q}q$ component of the $\eta$
mesons is introduced explicitely, providing
a contribution, $B_{2}^{g}$, which is 
an important part of the final result. In absence of gluons, the
symmetry $U(1)$ is not broken. In our case, the $U(1)$
symmetry is broken through the 't Hooft interaction term \cite{'tHooft:1986nc}%
\ introduced in the Lagrangian (\ref{A.01}), 
making our results consistent with those
of refs. \cite{Agaev:2010zz,Agaev:2003kb}. 

As stated at the end of the previous section, once the $\eta DA$ has been
obtained at the scale $Q_{0}^{2}$ and evolved to $Q^{2}$ according to Eqs.
(\ref{pDA0eta})-(\ref{pDA1eta}), the only remaining unknown for the evaluation
of the $\eta TFF$ according to Eq. (\ref{tff_R_T3}) is the constant $C_{3}$ of
the higher twist term. To this aim, three different scenarios have been
considered, corresponding to a contribution from this term to the form factor
at $Q_{0}^{2}=1%
\operatorname{GeV}%
^{2}$ of 10\% $(C_{3}=1.02\cdot10^{-2}%
\operatorname{GeV}%
^{3})$, 20\% $(C_{3}=2.04\cdot10^{-2}%
\operatorname{GeV}%
^{3})$\ and 30\% $(C_{3}=3.06\cdot10^{-2}%
\operatorname{GeV}%
^{3})$. The cutoff parameter $M$ varies between 487 MeV, for $C_{3}%
=1.02\cdot10^{-2}%
\operatorname{GeV}%
^{3},$ 557 Mev, for $C_{3}=2.04\cdot10^{-2}%
\operatorname{GeV}%
^{3},$ and 638 MeV, $C_{3}=3.06\cdot10^{-2}%
\operatorname{GeV}%
^{3}.$ We show in Fig \ref{Q01_0} the obtained result for $\eta TFF$
and in Fig. \ref{Q01_1} a detail of the region between $Q^{2}=0$ and $Q^{2}=10%
\operatorname{GeV}%
^{2}.$

The results for the $\eta TFF$, shown in Figs. \ref{Q01_0} and \ref{Q01_1},
exhibit a very good description of the experimental data in the whole
kinematic region. For completeness we have included in the figures
the $C_{3}=0$ case and the asymptotic vale for the $\eta TFF,$ $Q^{2}%
F_{\gamma\gamma^{\ast}\eta}\left(  Q^{2}\right)  =(5\sqrt{2}f_{\eta}%
^{\,q}+2f_{\eta}^{\,s})=0.181%
\operatorname{GeV}%
.$ It may be interesting to notice that the value of the asymptotic $\eta
TFF$\ is very close to the value of the asymptotic $\pi TFF,$ $Q^{2}%
F_{\gamma\gamma^{\ast}\pi}\left(  Q^{2}\right)  =\sqrt{2}f_{\pi}=0.185%
\operatorname{GeV}%
.$ It is also clear that, 
at variance with what happens in the 
$ \pi $ case, 
to explain the eta data some
$C_{3}$ contribution may be needed only in the region around
$Q^{2}=20-30%
\operatorname{GeV}%
^{2}$.
{Anyway, a complete discussion is obtained only}
by comparing the present results for
the $\pi TFF$ with those of Ref. \cite{Noguera:2010fe}. In the $\pi$
case, the $\ C_{3}$ contribution was crucial to reproduce 
the data in the region $Q^{2}=10-20%
\operatorname{GeV}%
^{2}$ and the calculated $\pi TFF$ crossed the asymptotic curve quite early
(around $Q^{2}=10%
\operatorname{GeV}%
^{2})$ and with a significative slope. In the $\eta$ case, the situation is
less dramatic: { the higher twist term} improves the TFF description only slowly,
and the theoretical result crosses softly the asymptotic value around
$Q^{2}=40%
\operatorname{GeV}%
^{2}.$

Another interesting point is the stability of the parameters. In calculating
the $\eta TFF$, we have adopted a procedure independent from that
used in Ref. \cite{Noguera:2010fe}, namely, $C_{3}$ and $M$ have been fitted
using the $\eta$ data only. The parameters used in both calculations have been
$\Lambda_{QCD}=0.226%
\operatorname{GeV}%
$ and $Q_{0}=1%
\operatorname{GeV}%
.$ Otherwise, in Ref \cite{Noguera:2010fe}, a fully model independent
calculation was performed, choosing $\phi_{\pi}\left(  x\right)  =1$ on the
basis of chiral symmetry. Here one is forced to choose a model for the
description of the $\eta DA$ at $Q_{0}^{2},$ and $C_{3}$ and $M$ have been
fixed within this model, independently from the $\pi$ case. Despite of this,
the result obtained in the two calculations are quite consistent. Varying the
weight of { the higher twist term} from 10\% to 30\% produces a change in $C_{3}$
from $1.02\cdot10^{-2}%
\operatorname{GeV}%
^{3}$ to $3.06\cdot10^{-2}%
\operatorname{GeV}%
^{3}$ in the $\eta$ case, to be compared to a variation of $C_{3}$ from
$0.99\cdot10^{-2}%
\operatorname{GeV}%
^{3}$ to $2.98\cdot10^{-2}%
\operatorname{GeV}%
^{3}$ in the $\pi$ case. The agreement is impressive.

On the other hand side, we found for the mass parameter a wider
variation. In the $\eta$ case one gets $M=560\pm70%
\operatorname{MeV}%
$,taking into account the uncertainty in $C_{3},$ to be compared to
$M=620\pm70%
\operatorname{MeV}%
$ for the $\pi$ case. Despite of these differences, the results can be
considered perfectly consistent with each other. The difference in the central
value of $M$ could imply that, for the pion, a larger contribution from the
transverse momentum is expected with respect to that for the eta particle. It
is indeed what has been obtained in Refs. \cite{Kroll:2010bf,Wu:2011gf}. The
values of $M$ could be compared with the value of $\left\langle k_{\perp
}\right\rangle $ given by P. Kroll \cite{Kroll:2010bf}, $\left\langle
k_{\perp}\right\rangle \simeq710%
\operatorname{MeV}%
$ for the pion and $\left\langle k_{\perp}\right\rangle \simeq390-440%
\operatorname{MeV}%
$ for the eta. It can be also compared with the $\beta_{\pi}$ parameter used
by \cite{Wu:2011gf}, which is related with the width of the gaussian
distribution of transverse momentum used by these authors, with the values
$\beta_{\pi}=668%
\operatorname{MeV}%
$ for the $u$-quark and $\beta_{\pi}\simeq530%
\operatorname{MeV}%
$ for the $s$-quark.
{A comparison of our parameters, based on a quark-flavor decomposition
of the relevant quantities (DAs, decay constants), with those
used in Ref. \cite{Xiao:2005af},
obtained within a singlet-octet decomposition, is instead rather involved.
The spirit of the present calculation
and those of Refs. \cite{Xiao:2005af,Wu:2011gf}
are rather different.
In our calculation, the known QCD evolution of the $DA$ governs the 
$Q^2$ dependence of the
form factor. The same $Q^2$ dependence is obtained, in Refs.
\cite{Xiao:2005af,Wu:2011gf}, through the 
$k_T$ dependence assumed for the
light-cone wave function of the mesons.
It is therefore significant that the two approaches 
provide similar results, describing probably, using different
tools, a similar mechanism.}

In the present discussion, the result on $\gamma^{\ast}\left(  q\right)
\rightarrow P\gamma,$ reported by BaBar for a time like $q^{2}=112%
\operatorname{GeV}%
^{2},$ ${q}^{2}{F}_{\gamma\gamma^{\ast}\eta}\left(  q^{2}\right)
{=0.229\pm0.030\pm0.008%
\operatorname{GeV}%
}$, has not been included. The reason is that the kinematics and dynamics of
this process is different from the ones studied here. There is no
symmetry relating $F_{\gamma\gamma^{\ast}P}(q^{2})$ {at one point} $q^{2}$ to
$F_{\gamma\gamma^{\ast}P}(Q^{2})$ {in the point} $Q^{2}=-q^{2}$. {The
coincidence is in the asymptotic value, which has been predicted for this
process to be}\cite{Lepage:1980fj} $-q^{2}F_{\gamma\gamma^{\ast}P}\left(
q^{2}\right)  =\sqrt{2}f_{P}\,(1-5\,\alpha_{s}\left(  q^{2}\right)  /3\pi),$
{when the contribution coming from the }$\alpha_{s}\left(  q^{2}\right)  ${
term could be disregarded}. In the present scheme we obtain
$Q^{2}F_{\gamma\gamma^{\ast}P}(Q^{2})=0.19$ GeV at $Q^{2}=112$ GeV$^{2},$
which implies a very slowly growing behavior of the $TFF$ {even for these high
values of the virtuality}.

In closing this section, it is useful to list items that prevent from
using the same formalism for the description of the $\eta^{\prime}TFF$. First
of all, as it has been previously noted, the NJL does not include confinement.
Therefore, if one uses the same expression, Eq. (\ref{eta.05}), in
the $\eta^{\prime}$ case, an imaginary part will appear in the DA at some
value of $x$. Secondly, the $\eta^{\prime}$ is basically a singlet
state and it can mix strongly with the two gluons state or, later, with some
$c\bar{c}$ component. These ingredients are not included in the present formalism.

\section{Conclusions}

In this paper, the $\eta TFF$ has been discussed in a formalism which connects
a low energy description of the hadron involved with a high energy description
based on a QCD perturbative formulation. The two descriptions are matched at
some scale $Q_{0}^{2}$. The scheme has been applied to describe the parton and
generalized parton distributions with notable
success\cite{Traini:1997jz,Scopetta:1997wk,Scopetta:1998sg,
Theussl:2002xp,Courtoy:2008nf} and, in particular, to the $\pi TFF$ in
\cite{Noguera:2010fe}. The formalism selects therefore two regions of
virtuality, separated at $Q_{0}^{2}.$ For $Q^{2}<Q_{0}^{2}$, use has been made
of the experimental parametrization of the $\eta TFF$ data. This has been done
to avoid model dependence in this region. At $Q^{2}>Q_{0}^{2}$, use has been
made of a high virtuality description, which incorporates the following
important physical ingredients: \textit{i)} a $\eta DA$ obtained in the NJL
model ; \textit{ii)} a mass cut-off in the definition of the $\eta TFF$ from
the $\eta DA,$ $M,$ \cite{Radyushkin:2009zg} which, interpreted from the point
of view of constituent models, takes into account the constituent mass,
transverse momentum effects and also higher twist effects; \textit{iii)} an
additional { higher twist term} into the definition of the $\eta TFF$ in the high
virtuality description, parameterized by a unique constant, $C_{3}$;
\textit{iv)} the two descriptions have to match at a virtuality $Q_{0}^{2}$, a
scale which is universal and should be the same for all observables.

In section II it has been shown that the dominant, twist two, expression for
the pseudoscalar $TFF,$ given in Eq. (\ref{F.28}) has to be corrected, for
including higher twist effects. The minimal correction would be the one given
in Eq. (\ref{tff_R_T3}).

The $\eta DA$ at $Q_{0}^{2}$ has been obtained in the NJL model. For that, the
parameters have been adjusted for a good reproduction of the $\eta$ sector 
with
the Pauli-Villars regularization. The obtained fits 
represent an
overall good description of the strange sector, not only for the masses, but
also for the meson decay constants. It is worth to strees that, in going from
SU(2) to SU(3), the number of parameters is increased by two, while the number
of new physical quantities, included in Table \ref{TableI}, are seven.
The obtained DAs in this model show consystency with other
analyses,where the DAs are parametrized 
\cite{Kroll:2010bf,Agaev:2010zz,Wu:2011gf}. 

Using $Q_{0}=1%
\operatorname{GeV}%
$ as matching point, the higher virtuality results of the $\eta TFF$ are well
reproduced. The $C_{3}$ term turns out to be relatively small. Its effect is to
reduce the value of the contribution to the twist two $\eta TFF$ only for
$Q^{2}<5$ GeV$^{2}$. Its value, $C_{3}=2.04\cdot10^{-2}%
\operatorname{GeV}%
^{3}$ for a 20\% of { higher twist} contamination at $Q_{0}^{2},$ is in 
perfect agreement with the one obtained for the $\pi TFF$ in Ref.
\cite{Noguera:2010fe}, $C_{3}=1.98\cdot10^{-2}%
\operatorname{GeV}%
^{3}.$ Moreover, the results are very stable with respect to variations of
this parameter.

The value obtained for $M=560%
\operatorname{MeV}%
$ is comparable with that of the pion case in Ref
\cite{Noguera:2010fe}, $M=620%
\operatorname{MeV}%
.$ The relative high value of $M$ in both cases can be
understood thinking that it includes the constituent quark mass, the
mean value of the transverse quark momentum and other higher twist
contributions. In turn, the higher value of $M$ for the $\pi$ than for the
$\eta$ can be related to the fact that the contribution of the quark
transverse momentum in the $\pi$ case is expected to be more important with
respect to the $\eta$ case \cite{Kroll:2010bf,Wu:2011gf}$.$

The calculation proves that all the BaBar results can be accommodated in the
present scheme, which only uses standard QCD ingredients and low virtuality
data. It must be emphasized that, in order to have a good description for both
$\pi$ and $\eta$, higher twist effects are important, as the modification from
Eq. (\ref{F.28}) to Eq. (\ref{tff_R_T3}) signals. It must be also noted that
the matching scale is as high as $1$ GeV, a feature already found in the
description of parton distributions when precision was to be attained. With
these ingredients, the calculation shows an excellent agreement with the data.

Let us conclude by stressing that we have justified the formalism developed in
Ref. \cite{Noguera:2010fe}\ to describe the $\pi TFF$ and we have extended it
to the $\eta TFF.$ The idea of the approach is that one can use models or
effective theories to describe the non perturbative sector, and QCD to
describe the perturbative one. In here, we have preferred to use data for the
low virtuality sector to avoid model dependence, but in building the $\eta DA$
at $Q_{0}^{2}$ we have used the NJL model. Higher twist effects (parametrized
in our case by $M$ and $C_{3}$) are small but crucial in order to attain an
excellent description of the $\pi$ and $\eta$ experimental results.

\appendix

\section{The $SU(3)$ NJL model for pseudoscalars mesons.}

\label{AppendixA}

In calculating the $\eta DA$, the minimal extension of the NJL model for
describing pseudoscalar mesons in SU(3), proposed in Ref \cite{Bernard:1987sg}%
, has been used:%
\begin{align}
\mathcal{L}  &  =\bar{q}\left(  x\right)  \left(  i\rlap{$/$}\hspace
*{-0.05cm}\partial-\mu\right)  q\left(  x\right)  +G\sum_{a=0}^{8}\left[
\left(  \bar{q}\lambda^{a}q\right)  ^{2}+\left(  \bar{q}i\gamma_{5}\lambda
^{a}q\right)  ^{2}\right]  -\nonumber\\
&  -K\,\left[  \det\left(  \bar{q}\left(  1+\gamma_{5}\right)  q\right)
+\det\left(  \bar{q}\left(  1-\gamma_{5}\right)  q\right)  \right]  \quad,
\label{A.01}%
\end{align}
where $\mu=\mathrm{diag}[\mu_{u},\mu_{d},\mu_{s}]$ is the matrix of the
current quark masses and $\lambda^{a},$ $a=0,...,8,$ are the SU(3) generators.
SU(2) will be considered a good symmetry, and, therefore, $\mu_{u}=\mu_{d}.$
As it is well known, the first consequence of the scalar interaction term is
to provide the constituent quark masses, $m_{u}=m_{d},\ m_{s},$ different from
the current ones. The main results are summarized here, while the reader is
referred to the section IV-B of Ref. \cite{Klevansky:1992qe} for details.

By defining the integrals:%
\begin{align}
I_{1}\left(  m\right)   &  =i\int\frac{d^{4}p}{\left(  2\pi\right)  ^{4}}%
\frac{1}{\left(  p^{2}-m^{2}+i\epsilon\right)  }~,\nonumber\\
I_{2}\left(  m_{i},m_{j},q^{2}\right)   &  =i\int\frac{d^{4}p}{\left(
2\pi\right)  ^{4}}\frac{1}{\left(  p^{2}-m_{i}^{2}+i\epsilon\right)  \left[
\left(  p-q\right)  ^{2}-m_{j}^{2}+i\epsilon\right]  }~, \label{A.02}%
\end{align}
the constituent masses are given by%
\begin{align}
m_{u}  &  =\mu_{u}+16\,G\,N_{c}\,m_{u}\,I_{1}\left(  m_{u}\right)
+32\,K\,N_{c}^{2}\,m_{u}\,I_{1}\left(  m_{u}\right)  \,m_{s}\,I_{1}\left(
m_{s}\right)  ~,\nonumber\\
m_{s}  &  =\mu_{s}+16\,G\,N_{c}\,m_{s}\,I_{1}\left(  m_{s}\right)
+32\,K\,N_{c}^{2}\,\left[  m_{u}\,I_{1}\left(  m_{u}\right)  \right]  ^{2}~,
\label{A.03}%
\end{align}
where $N_{c}$ is the number of colors. The vacuum expectation values for the
condesates of the quarks of flavor $q_{i}$ are%
\begin{equation}
\left\langle \bar{q}_{i}q_{i}\right\rangle =4\,m_{i}\,I_{1}\left(
m_{i}\right)  -4\,\mu_{i}\,I_{1}\left(  \mu_{i}\right)  ~, \label{A.05}%
\end{equation}
where the expectation value of the $\bar{q}q$ in the perturbative vacuum has
been substracted from the expectation value in the true vacuum. The last term
in Eq. (\ref{A.05}) is negligible in the $u-d$ quark sector, while it becomes
important in the strange sector.

The next step is the description of the pseudoscalar states. For the pion and
kaon case, by defining the quantities%
\begin{align}
K_{3}  &  =G+2\,N_{c}\,K\,m_{s}\,I_{1}\left(  m_{s}\right)  ~,\nonumber\\
K_{6}  &  =G+2\,N_{c}\,K\,m_{u}\,I_{1}\left(  m_{u}\right)  ~,
\end{align}
and
\begin{equation}
D\left(  m_{i},m_{j},q^{2}\right)  =I_{1}\left(  m_{i}\right)  +I_{1}\left(
m_{j}\right)  +\left(  \left(  m_{i}-m_{j}\right)  ^{2}-q^{2}\right)
\,I_{2}\left(  m_{i},m_{j},q^{2}\right)  ~,
\end{equation}
one has that the pion and kaon masses are obtained solving the equations:%
\begin{align}
1-8\,N_{c}\,K_{3}\,D\left(  m_{u},m_{u},m_{\pi}^{2}\right)   &
=0~,\nonumber\\
1-8\,N_{c}\,K_{6}\,D\left(  m_{u},m_{s},m_{K}^{2}\right)   &  =0~.
\end{align}
The couplings of the pion and the kaon to the quarks are given by:%
\[
g_{\pi qq}^{2}=\left[  4\,N_{c}\,\frac{d}{dq^{2}}D\left(  m_{u},m_{u}%
,q^{2}\right)  \right]  _{q^{2}=m_{\pi}^{2}}^{-1}~,
\]%
\begin{equation}
g_{Kqq}^{2}=\left[  4\,N_{c}\,\frac{d}{dq^{2}}D\left(  m_{u},m_{s}%
,q^{2}\right)  \right]  _{q^{2}=m_{K}^{2}}^{-1}~,
\end{equation}
and the decay constants are:%
\begin{equation}
F_{\pi}=-4\,N_{c}\,g_{\pi qq}\,m_{u}\,I_{2}\left(  m_{u},m_{u},m_{\pi}%
^{2}\right)  ~,
\end{equation}%
\begin{equation}
F_{K}=\frac{2\,N_{c}\,g_{Kqq}}{m_{K}^{2}}\,\left\{  \left(  m_{s}%
+m_{u}\right)  \,D\left(  m_{u},m_{s},m_{K}^{2}\right)  -2\,m_{u}%
\,I_{1}\left(  m_{u}\right)  -2\,m_{s}\,I_{1}\left(  m_{s}\right)  \right\}
~,
\end{equation}
where $F_{\pi,K}=f_{\pi,K}/\sqrt{2}$.

The $\eta$ particle deserves a more careful discussion, due to its mixing with
the $\eta^{\prime}$ particle. Working in the flavor basis, one can define
\begin{align}
K_{q}  &  =G-2\,N_{c}\,K\,m_{s}\,I_{1}\left(  m_{s}\right)  ~,\nonumber\\
K_{qs}  &  =-2\sqrt{2}\,N_{c}\,K\,m_{u}\,I_{1}\left(  m_{u}\right)  ~,\\
{\mathcal{K}}  &  =K_{q}\,G-K_{qs}^{2}~.\nonumber
\end{align}
The interaction in the $\eta-\eta^{\prime}$ sector can be described by the
expression%
\begin{equation}
\left(  i\gamma_{5}\lambda^{i}\right)  \left[  M_{ij}\right]  \left(
i\gamma_{5}\lambda^{j}\right)
\end{equation}
with $i,j=q,s$, $\lambda^{q}=\mathrm{diag}[1,1,0]$, $\lambda^{s}%
=\mathrm{diag}[0,0,\sqrt{2}]$ and the interaction matrix is given by%
\begin{equation}
\left[  M_{ij}\right]  =\frac{1}{D_{\eta}}\left(
\begin{array}
[c]{cc}%
a_{qq} & a_{qs}\\
a_{qs} & a_{ss}%
\end{array}
\right)  ~,
\end{equation}
with%
\begin{align}
a_{qq}  &  =2\left(  K_{q}-8\,{\mathcal{K}} \,N_{c}\,D\left(  m_{s}%
,m_{s},q^{2}\right)  \right)  ~,\nonumber\\
a_{ss}  &  =2\left(  G-8\,{\mathcal{K}} \,N_{c}\,D\left(  m_{u},m_{u}%
,q^{2}\right)  \right)  ~,\nonumber\\
a_{qs}  &  =2 \, K_{qs} ~,\\
D_{\eta}\left(  q^{2}\right)   &  = (a_{qq}a_{ss} -a_{qs}^{2})/(4
{\mathcal{K}})~.\nonumber
\end{align}
The $\eta$ mass is obtained solving the equation%
\begin{equation}
D_{\eta}\left(  m_{\eta}^{2}\right)  =0~. \label{A.20}%
\end{equation}
In a neighborhood of $q^{2}=m_{\eta}^{2}$ the interaction can be written as%
\begin{align}
&  \left(  -\sin\phi\,\lambda^{s}+\cos\phi\,\lambda^{q}\right)  \frac{-g_{\eta
qq}}{q^{2}-m_{\eta}^{2}}\left(  -\sin\phi\,\lambda^{s}+\cos\phi\,\lambda
^{q}\right) \nonumber\\
&  =\frac{a_{qq}}{D_{\eta}}\left(  \epsilon_{\eta}\,\lambda^{s}+\lambda
^{q}\right)  \left(  \epsilon_{\eta}\,\lambda^{s}+\lambda^{q}\right)  ~,
\label{A.22}%
\end{align}
with $\epsilon_{\eta}=a_{qs}/a_{qq}.$ In obtaining the right hand side of this
equation, use has been made of Eq. (\ref{A.20}), which implies $a_{ss}%
=a_{qs}^{2}/a_{qq}.$ From (\ref{A.22}) one has%
\[
\cos\phi=\frac{1}{\sqrt{1+\epsilon_{\eta}^{2}}}\ \ ,\ \ \ \ \ \ \ \ \ \ \ \sin
\phi=\frac{-\epsilon_{\eta}}{\sqrt{1+\epsilon_{\eta}^{2}}}~,
\]%
\begin{equation}
g_{\eta qq}^{2}=-\left.  \frac{\left(  1+\epsilon_{\eta}^{2}\right)  \,a_{qq}%
}{dD_{\eta}/dq^{2}}\right\vert _{q^{2}=m_{\eta}^{2}}~.
\end{equation}
For the flavor decay constants, one has%
\begin{align}
F_{\eta}^{q}  &  =-12\,g_{\eta qq}\,\cos\phi\,m_{u}\,I_{2}\left(  m_{u}%
,m_{u},m_{\eta}^{2}\right)  ~,\nonumber\\
F_{\eta}^{s}  &  =12\,g_{\eta qq}\,\sin\phi\,m_{s}\,I_{2}\left(  m_{s}%
,m_{s},m_{\eta}^{2}\right)  ~, \label{Af}%
\end{align}
where $F_{\eta}^{q,s}=f_{\eta}^{q,s}/\sqrt{2}$.

We need to evaluate the integrals defined in Eq. (\ref{A.02}). Due
to the point-like character of the interaction, the lagrangian Eq.
(\ref{A.01}) is not renormalizable and a regularization procedure for
these integrals 
has to be defined. We use the Pauli-Villars regularization in order
to render the occurring integrals finite. This means that, for integrals like
the ones defined in Eq.~(\ref{A.02}), we make the following
replacement,
\begin{equation}
I_{1}\left(  m_{i}\right)  \longrightarrow\sum_{\ell=0}%
^{2}c_{\ell}\,I_{1}\left(  M_{\ell,i}\right)
\ \ \ \ \ \ \ \ I_{2}\left(  m_{i},m_{j},q^{2}\right)  \longrightarrow
\sum_{\ell=0}^{2}c_{\ell}\,I_{2}\left(  M_{\ell,i},M_{\ell,j}%
,q^{2}\right)
\end{equation}
with $M_{\ell,j}^{2}=m_{j}^{2}+\ell\,\Lambda^{2}$, $c_{0}=c_{2}=1,$
$c_{1}=-2.$ Here, for simplicity, we choose the same $\Lambda$ value
for the strange and the nonstrange sector. According to these prescriptions
one finds%
\begin{equation}
I_{1}\left(  m_{i}\right)  =\frac{1}{16\,\pi^{2}}\left[  -2\,M_{1,i}^{2}%
\ln\frac{M_{1,i}^{2}}{m_{i}^{2}}+M_{2,i}^{2}\ln\frac{M_{2,i}^{2}}{m_{i}^{2}%
}\right]  ~,
\end{equation}%
\begin{equation}
I_{2}\left(  m_{i},m_{j},q^{2}\right)  =\frac{1}{32\pi^{2}}\sum_{\ell=0}%
^{2}c_{\ell}\left[  \left(  \ln\frac{M_{1,i}^{2}}{m_{i}^{2}}+\ln\frac
{M_{1,j}^{2}}{m_{j}^{2}}\right)  +\frac{M_{1,j}^{2}-M_{1,i}^{2}}{q^{2}}%
\ln\frac{M_{\ell,j}^{2}}{M_{\ell,i}^{2}}+\Phi_{\ell}\right]  ~, \label{A.25}%
\end{equation}
with%
\begin{equation}
\Phi_{\ell}=\frac{2}{q^{2}}\sqrt{-\lambda\left(  M_{\ell,i}^{2},M_{\ell,j}%
^{2},q^{2}\right)  }\left[  \arctan\frac{q^{2}+M_{1,j}^{2}-M_{1,i}^{2}}%
{\sqrt{-\lambda\left(  M_{\ell,i}^{2},M_{\ell,j}^{2},q^{2}\right)  }}%
+\arctan\frac{q^{2}-M_{1,j}^{2}+M_{1,i}^{2}}{\sqrt{-\lambda\left(  M_{\ell
,i}^{2},M_{\ell,j}^{2},q^{2}\right)  }}\right]
\end{equation}
where $\lambda\left(  M_{\ell,i}^{2},M_{\ell,j}^{2},q^{2}\right)  $ is the
K\"{a}ll\'{e}n lambda.

Now, we fix the parameters of the model. Looking at the lagrangian,
we have a five parameters model, $\mu_{u},$ $\mu_{s},$ $G,$ $K$ and $\Lambda.$
Nevertheless, it is more intuitive to organize the fit of the parameters in
terms of $\mu_{u},$ $\mu_{s},$ $m_{u},$ $m_{s}$ and $\Lambda,$ using equations
(\ref{A.03}) to determine $G$ and $K.$ We impose $m_{u}=275%
\operatorname{MeV}%
,$ in order to have $m_{\eta}^{\exp}<2m_{u}.$ Then, $\mu_{u}$ and $\Lambda$
are obtained in recovering the values of $F_{\pi}$ and $m_{\pi}.$ At this step
one has%
\[
\mu_{u}=6.69%
\operatorname{MeV}%
,\ \ m_{u}=275%
\operatorname{MeV}%
,\ \ \Lambda=740%
\operatorname{MeV}%
\ \ \longrightarrow\ \ \ m_{\pi}=138%
\operatorname{MeV}%
,\ \ F_{\pi}=92.2%
\operatorname{MeV}%
,\ \ \left\langle \bar{u}u\right\rangle =\left(  -227%
\operatorname{MeV}%
\right)  ^{3}%
\]
determining the SU(2) sector. Then, $\mu_{s}$ and $m_{s}$ have been fixed by
requiring a good overall fit of masses ($m_{K},$ $m_{\eta},$ $m_{\eta^{\prime
}})$ and decay constants ($F_{K},$ $F_{\eta}^{q},$ $F_{\eta}^{s}).$ In table
\ref{TableI} two different sets of parameters are given, together with the
obtained results. Using set I, by imposing $m_{\eta}\leq2m_{u},$ a good
agreement for the masses and a slightly less good agreement for the $F_{\eta
}^{q,s}$ is obtained. On the other hand, in set II a very good agreement for
$F_{\eta}^{q,s}$ is obtained, with a slightly worse result for the masses.

In the light-front calculation one needs the integral%
\begin{align}
\tilde{I}_{2}\left(  x,m_{i},m_{P}^{2}\right)   &  =i\int\frac{d^{4}k}%
{(2\pi)^{4}}\,\,\frac{\delta\left(  x-1+\frac{k^{+}}{P^{+}}\right)  }{\left[
(k-P)^{2}-m_{i}^{2}+i\epsilon\right]  \left(  k^{2}-m_{i}^{2}+i\epsilon
\right)  }\nonumber\\
&  =-\theta(x)\theta(1-x)\,\frac{1}{(4\pi)^{2}}\,\sum_{\ell=1}^{2}c_{\ell
}\,\ln\,\frac{m_{i}^{2}-(1-x)\,x\,m_{P}^{2}}{M_{\ell,i}^{2}-(1-x)\,x\,m_{P}%
^{2}}\qquad. \label{A.28}%
\end{align}
Clearly, all one needs to calculate the $\eta DA$ are the $\eta$ and quark
masses and the value of the cutoff parameter. For the DA calculations, the
values $m_{u}=275%
\operatorname{MeV}%
,$ $m_{s}=435%
\operatorname{MeV}%
,$ $\Lambda=740%
\operatorname{MeV}%
$ and $m_{\eta}=548%
\operatorname{MeV}%
$ have been chosen.

\section*{Acknowledgements}

This work was supported in part by HadronPhysics3, a
FP7-Infrastructures-2011-1 Program of the European Commission under Grant
283288, by the MICINN (Spain) grant FPA2010-21750-C02-01, by Generalitat
Valenciana, grant Prometeo2009/129, and by ``Partonic structure of nucleons,
mesons and light nuclei'', an INFN (Italy, Perugia) - MICINN (Spain, Valencia)
exchange agreement.

\end{document}